\documentclass[twocolumn,english]{revtex4-1}
\usepackage[T1]{fontenc}
\usepackage[latin9]{inputenc}
\usepackage{color}
\usepackage{amsmath}
\usepackage{graphicx}
\usepackage{esint}

\makeatletter
\usepackage{babel}
\usepackage{amsfonts}

\makeatother

\usepackage{babel}
\begin{document}
\title{Elastic gauge fields and zero-field 3D quantum Hall effect in hyperhoneycomb
lattices }
\author{Sang Wook Kim and Bruno Uchoa}
\affiliation{Department of Physics and astronomy, University of Oklahoma, Norman,
Oklahoma 73019, USA }
\date{\today}
\begin{abstract}
Dirac materials respond to lattice deformations as if the electrons
were coupled to gauge fields. We derive the elastic gauge fields in
the hyperhoneycomb lattice, a three dimensional (3D) structure with
trigonally connected sites. In its semimetallic form, this lattice
is a nodal-line semimetal with a closed loop of Dirac nodes. Using
strain engineering, we find a whole family of strain deformations
that create uniform nearly flat Landau levels in 3D. We propose that
those Landau levels can be created and tuned in metamaterials with
the application of a simple uniaxial temperature gradient. In the
3D quantum anomalous Hall phase, which is topological, we show that
the components of the elastic Hall viscosity tensor are multiples
of $\eta_{H}=\beta^{2}\sqrt{3}/\left(8\pi a^{3}\right)$, where $\beta$
is an elastic parameter and $a$ is the lattice constant. 
\end{abstract}
\maketitle
\emph{Introduction. }In honeycomb lattices such as graphene \cite{Neto},
strain deformations couple to electronic degrees of freedom as gauge
fields and can induce Landau level (LL) quantization with very large
effective pseudomagnetic fields \cite{Levy,Gomes,Rechtsman,Guinea}.\emph{
}When the chemical potential is inside the gap of the LLs, the Hall
conductivity per valley is quantized and the system is expected to
show a zero-field quantum Hall effect (QHE). Due to the dispersion
of the LLs, Hall conductivity quantization is not common in three
dimensions (3D), and may occur only in extremely anisotropic systems
such as Bechgaard salts \cite{Balicas,McKerman}, Bernal graphite
\cite{Bernevig,Arovas}, and in nodal-line semimetals \cite{Mullen,Lim,Rhim}.
Even in strongly anisotropic systems such as in nodal line semimetals,
the physical implementation of the 3D QHE is challenging due to the
unusual toroidal field geometry required \cite{Mullen}. With the
help of strain engineering, one may in principle design 3D LLs with
well defined gaps in between from real space configurations of magnetic
field that would be otherwise impractical to realize. 

In this Rapid communication, we derive the elastic gauge fields that
follow from arbitrary lattice deformations in the hyperhoneycomb lattice,
a natural 3D generalization of the honeycomb geometry where all sites
are connected by coplanar trigonal bonds, as shown in Fig. 1a. In
the semimetallic form, this lattice is an example of a nodal-line
semimetal \cite{Mullen,kane,Weng,Yu,Heikkila,Chen,Xie}. We identify
a whole family of lattice deformations that produce uniform nearly
flat LLs in 3D, a prerequisite for the 3D zero-field QHE. We show
that this family of non-trivial deformations can be physically implemented
with the application of a simple temperature gradient along the axis
perpendicular to the nodal line, leading to a tunable metal-insulator
transition in the bulk. The strain deformations can be uniquely specified
by the set of thermal expansion coefficients of the crystal. We propose
that a tunable temperature controlled 3D zero-field QHE can be implemented
in acoustic metamaterials \cite{zhu-1}.

In the presence of topological states, the topological invariants
can manifest in the elastic response of the crystal through phonons.
In the 3D quantum anomalous Hall (QAH) phase \cite{Kim}, which is
the extension of the Haldane model \cite{Haldane} to the hyperhoneycomb
lattice, we also calculate the elastic Hall viscosity tensor $\eta_{\mu\nu\rho\gamma}$.
Also known as the phonon Hall viscosity \cite{Barkeshli}, this quantity
is analogous to the dissipationless viscous response of electrons
in the quantum Hall regime \cite{Avron1,Avron2,Read} and is topological
in nature. We show that the components of the Hall viscosity tensor
are $\pm\eta_{H}$ or $\pm2\eta_{H}$ (or zero), with $\eta_{H}=\beta^{2}\sqrt{3}/\left(8\pi a^{3}\right)$,
where $\beta$ is an elastic parameter and $a$ is the lattice constant. 

\begin{figure}[b]
\begin{centering}
\includegraphics[scale=0.32]{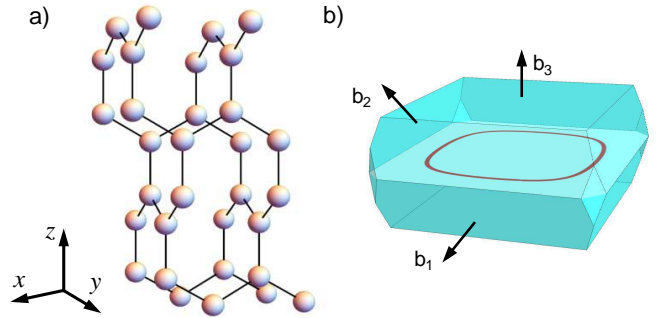}
\par\end{centering}
\caption{{\small{}a) Hyperhoneycomb lattice, with four atoms per unit cell.
All sites are linked by coplanar trigonal bonds spaced by 120$^{\circ}$.
b) Brillouin zone (BZ) of the hyperhoneycomb lattice, with the nodal
line shown in red. The arrows show the reciprocal lattice vectors. }}
\end{figure}

\emph{Hamiltonian. }The hyperhoneycomb lattice has four sites per
unit cell $\mu=1,\ldots,4$ and is generated by the lattice vectors
$\mathbf{a}_{1}=(\sqrt{3},0,0)$, $\mathbf{a}_{2}=(0,\sqrt{3},0)$,
and $\mathbf{a}_{3}=(-\sqrt{3}/2,\sqrt{3}/2,3)$, in units of the
lattice constant $a$. In the momentum space, the reciprocal lattice
is generated by the vectors $\mathbf{b}_{1}=(2\pi/\sqrt{3},0,-\pi/3)$,
$\mathbf{b}_{2}=(0,-2\pi/\sqrt{3},\pi/3)$ and $\mathbf{b}_{3}=(0,0,2\pi/3)$,
shown in Fig. 1b. The tight-binding Hamiltonian is a $4\times4$ matrix
\cite{Mullen}
\begin{equation}
\mathcal{H}_{0,\mu\nu}(\mathbf{k})=-t_{0}\sum_{\vec{\delta}_{\mu\nu}}\text{e}^{i\mathbf{k}\cdot\vec{\delta}_{\mu\nu}},\label{eq:1}
\end{equation}
where $t_{0}$ is the hopping amplitude, $\vec{\delta}_{\mu\nu}$
are the nearest neighbor (NN) vectors between sites of species $\mu$
and $\nu$ and $\mathbf{k}$ is the momentum measured from the center
of the Brillouin zone (BZ). In total, there are six NN vectors $\vec{\delta}_{12}=\left(\pm\sqrt{3}a/2,0,a/2\right)$,
$\vec{\delta}_{34}=\left(0,\pm\sqrt{3}a/2,a/2\right)$, $\vec{\delta}_{14}=\left(0,0,-a\right)$
and $\vec{\delta}_{23}=\left(0,\,0,a\right)$. The low energy bands
of this lattice have a line of Dirac nodes $\mathbf{k}_{0}=[k_{x}(s),k_{y}(s),0]$
in the $k_{z}=0$ plane, which can be written in terms of some parameter
$s$ that satisfies the equation $4\cos[3k_{x}(s)/2]\cos[3k_{y}(s)/2]=1$.
The low energy projected Hamiltonian is described by a $2\times2$
matrix expanded around the nodal line 
\begin{equation}
\mathcal{H}_{0,p}(\mathbf{q})=\left[v_{x}(s)q_{x}+v_{y}(s)q_{y}\right]\sigma_{1}+v_{z}(s)q_{z}\sigma_{2}
\end{equation}
where $\mathbf{q}\equiv\mathbf{k}-\mathbf{k}_{0}(s)$ is the relative
momentum, $\sigma_{1},\:\sigma_{2}$ are the two off-diagonal Pauli
matrices and
\begin{align}
v_{x}(s) & =\frac{\sqrt{3}}{1+\alpha^{2}}\sin\left(\frac{\sqrt{3}}{2}k_{x}(s)\right)t_{0}\nonumber \\
v_{y}(s) & =\frac{\alpha^{2}\sqrt{3}}{1+\alpha^{2}}\sin\left(\frac{\sqrt{3}}{2}k_{y}(s)\right)t_{0}\\
v_{z}(s) & =-\frac{3\alpha}{1+\alpha^{2}}t_{0},\nonumber 
\end{align}
are the velocities of the quasiparticles, with $\alpha(s)\equiv2\cos[\sqrt{3}k_{x}(s)a/2]$
\cite{Kim}. The energy spectrum of the quasiparticles is $E_{0}(\mathbf{q})=\pm\sqrt{(v_{x}q_{x}+v_{y}q_{y})^{2}+v_{z}^{2}q_{z}^{2}}.$
The wavefunctions have a $\pi$ Berry phase for closed line trajectories
that encircle the nodal loop. 

\emph{Elastic gauge fields.} The inclusion of lattice deformations
can be done by locally changing the distance between lattice sites,
which affect value of the hopping constant. Expanding it to lowest
order in the displacement of the lattice, 
\begin{equation}
t\left(\vec{\delta}^{(n)}+\delta\mathbf{r}\right)\approx t_{0}+\frac{\beta}{a^{2}}\delta_{i}^{(n)}\delta_{j}^{(n)}u_{ij}+\mathcal{O}\left(\delta r^{2}\right),\label{t2}
\end{equation}
with $n=1,\ldots,6$ indexing the 6 NN lattice vectors $\vec{\delta}^{(n)}$,
$u_{ij}=\frac{1}{2}\left(\partial_{i}u_{j}+\partial_{j}u_{i}\right)$
is the strain tensor defined in terms of the displacement field $\mathbf{u}$
of the lattice and $\beta=a\frac{\partial t}{\partial r}=\frac{\partial\log t}{\partial\log r}$
is the Grüneisen parameter of the model \cite{note1}. Including the
lattice distortions in Hamiltonian (\ref{eq:el}), one gets two terms,
$\mathcal{H}_{p}=\mathcal{H}_{0,p}+\mathcal{H}_{el}$, where
\begin{equation}
\mathcal{H}_{el}=\frac{3}{4}\frac{\beta}{a}v_{z}\left(u_{xx}+u_{yy}-2u_{zz}\right)\sigma_{1}-\frac{\beta}{a}\left(v_{x}u_{xz}+v_{y}u_{yz}\right)\sigma_{2}\label{eq:el}
\end{equation}
is the elastic contribution. As in the 2D case (graphene), the deformation
of the lattice couples to the Dirac fermions as an elastic gauge field
$\mathbf{A}$. It is convenient to rewrite the Hamiltonian in the
more familiar form 
\begin{equation}
\mathcal{H}_{p}(\mathbf{q})=\left[v_{x}\left(q_{x}+A_{x}\right)+v_{y}\left(q_{y}+A_{y}\right)\right]\sigma_{1}+v_{z}\left(q_{z}+A_{z}\right)\sigma_{2},\label{Ho3}
\end{equation}
where 
\begin{align}
A_{x}(s) & =\frac{v_{x}v_{z}}{v_{\rho}^{2}}\frac{3\beta}{4a}\left(u_{xx}+u_{yy}-2u_{zz}\right)\nonumber \\
A_{y}(s) & =\frac{v_{y}v_{z}}{v_{\rho}^{2}}\frac{3\beta}{4a}\left(u_{xx}+u_{yy}-2u_{zz}\right)\label{Ay}\\
A_{z}(s) & =-\frac{\beta}{a}\left(\frac{v_{x}}{v_{z}}u_{xz}+\frac{v_{y}}{v_{z}}u_{yz}\right)\nonumber 
\end{align}
are the components of the elastic gauge field along the nodal line,
with $v_{\rho}^{2}(s)=v_{x}^{2}(s)+v_{y}^{2}(s)$. The definition
of the $A_{x}$ and $A_{y}$ components is to a degree arbitrary.
In (\ref{Ay}) we chose the most symmetric combination, although this
choice has no effect in physical observables. 

Those gauge fields can be associated to a pseudomagnetic field $\mathbf{B}=\nabla\times\mathbf{A}$,
which follows from lattice deformations and hence must preserve time
reversal symmetry (TRS). While pseudo magnetic fields couple to the
Dirac fermions similarly to conventional magnetic fields and can produce
Landau level (LL) quantization, they create a zero net magnetic flux
at each lattice site. Therefore, electrons sitting at opposite points
in the nodal line are related by TRS and must necessarily couple to
opposite $\mathbf{B}$ fields. In order to produce zero-field quantum
Hall effect, one needs to create 3D LL quantization with well defined
gaps in between. In 2D, the conventional Hall conductivity $\sigma_{xy}$
is a dimensionless and quantized in units of $e^{2}/h$. In 3D, it
has an extra unit of inverse length. According to Halperin \cite{Halperin},
the Hall conductivity tensor is $\sigma_{ij}=e^{2}/(2\pi h)\epsilon_{ijk}G_{k}$,
where \textbf{$\mathbf{G}$ }is a reciprocal lattice vector (and could
be zero). In general, a finite Hall conductivity in 2D (3D) is allowed
whenever the chemical potential is in the gap between different LLs,
and implies in the existence of chiral edge (surface) states. At zero
field, the Hall conductivity tensor due to pseudomagnetic fields does
not create chiral charge currents as in the conventional quantum Hall
effect, but rather a valley current.

\begin{figure}[t]
\begin{centering}
\includegraphics[scale=0.43]{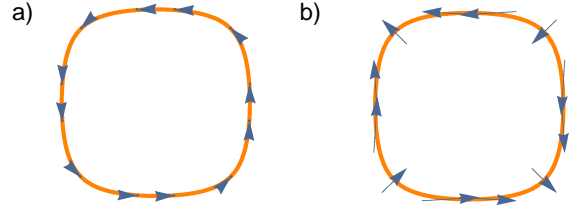}
\par\end{centering}
\caption{{\small{}Pseudomagnetic field $\mathbf{B}$ along the nodal line for
two different strain field configurations. a) $\mathbf{u}=(2xz,2yz,z^{2})$
and b) $\mathbf{u}=(2yz,2xz,0)$. Both configurations lead to uniform
$\mathbf{B}$ fields in real space, but only the former produces nearly
flat LLs. }}
\end{figure}

\begin{figure*}[t]
\begin{centering}
\includegraphics[scale=0.33]{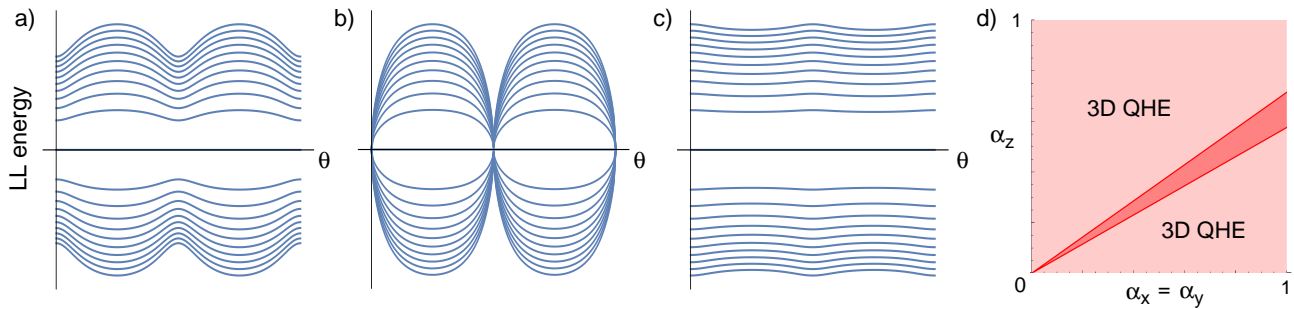}
\par\end{centering}
\caption{{\small{}Energy of the Landau levels (LLs) around the nodal line vs.
polar angle $\theta$, ($0\protect\leq\theta<\pi$) for three strain
configurations: (a) $\mathbf{u}=(2xz,2yz,z^{2})$, (b) $\mathbf{u}=(2yz,2xz,0)$
and (c) $\mathbf{u}=(2xz,2yz,0)$. The former and the latter configurations
belong to a broader family of deformations $\mathbf{u}=(\alpha_{x}xz,\alpha_{y}yz,\alpha_{z}z^{2})$
that produce nearly flat LLs in 3D. In (b), the LLs collapse at discrete
points of the nodal line, preventing the zero-field QHE. Application
of a uniform temperature gradient, $\Delta T\propto z$, creates strain
fields in that family. (d) Phase space for $\alpha_{x}=\alpha_{y}$
and $\alpha_{z}$ with a zero-field 3D QHE (light red regions).}}
\end{figure*}

\emph{Strain engineering. }\textcolor{black}{In all possible strain
configurations, the effective Hamiltonian (\ref{Ho3}) has the form
$\mathcal{H}_{p}(\mathbf{q})=h_{1}\sigma_{1}+h_{2}\sigma_{2}$. In
specific, for configuration}\textcolor{red}{{} }\textcolor{black}{$\mathbf{u}=(2xz,2yz,z^{2})$,
\begin{align}
h_{1} & =v_{x}q_{x}+v_{y}q_{y}\label{h1}\\
h_{2} & =v_{z}q_{z}-\frac{\beta}{a}v_{x}x-\frac{\beta}{a}v_{y}y.\label{h2}
\end{align}
The corresponding pseudomagnetic field $\mathbf{B}=(-v_{y}/v_{z},v_{x}/v_{z},0)$
forms a closed loop in the BZ around the nodal line, as shown in Fig.
2a. In order to calculate the spectrum of Landau levels, we generically
define the canonically conjugated ladder operators $a=\frac{1}{\omega}\left(h_{1}+ih_{2}\right)\text{ and }a^{\dagger}=\frac{1}{\omega}\left(h_{1}-ih_{2}\right)$,
which satisfy $[a,a^{\dagger}]=2i[h_{2},h_{1}]/\omega^{2}=1$. The
parameter 
\begin{equation}
\omega(s)=\sqrt{\frac{\beta}{a}}\left[2v_{x}^{2}(s)+2v_{y}^{2}(s)\right]^{\frac{1}{2}},\label{omega}
\end{equation}
}is the analog of the cyclotronic frequency. Taking the square of
the Hamiltonian, $\mathcal{H}_{0}^{2}=\omega^{2}\left[a^{\dagger}a+\frac{1}{2}\right]1_{2\times2}-\frac{1}{2}\omega^{2}\sigma_{3}$,
that results in the spectrum of LLs parametrized along the nodal line,
\begin{equation}
E_{N}(s)=\text{sgn}(N)\omega(s)\sqrt{|N|},\label{eq:LLs}
\end{equation}
with $N\in\mathbb{Z},$ as shown in Fig. 3a. The energy spectrum has
a zeroth LL, as expected for Dirac fermions \cite{Neto,Goerbig},
and a clear gap between the first few LLs. That permits the emergence
of a zero-field QHE due to strain whenever the chemical potential
lays in the LL gap. Even though there are many deformation sets producing
uniform pseudomagnetic fields in real space, not all of them create
3D LL quantization with well defined gaps in between. For the strain
configuration shown in Fig. 2b, $\mathbf{u}=(2yz,2xz,0)$, which corresponds
to the pseudomagnetic field $\mathbf{B}=(-v_{x}/v_{z},v_{y}/v_{z},0)$,
the parameter $\omega(s)=\sqrt{(\beta/a)|v_{x}(s)v_{y}(s)|}$ has
zeros along the nodal line (see Fig. 3b), where all LLs collapse.
In that configuration, although the LLs are well defined away from
those points, their dispersion does not lead to a well defined gap
in the excitation spectrum, and hence the system does not have a zero-field
QHE.

In general, one can define families of strain deformations that lead
to a 3D zero-field QHE. While the energy spectrum is generically defined
by Eq. (\ref{eq:LLs}), in those families $\omega(s)=\sqrt{2|[h_{2},h_{1}]|}$
can be non zero for all points along the nodal line. For instance,
one can build a family of strain deformations 
\begin{equation}
\mathbf{u}=(\alpha_{x}xz,\alpha_{y}yz,\alpha_{z}z^{2}),\label{eq:u-1}
\end{equation}
where the constants $\alpha_{i}$ ($i=x,y,z)$ are such that $\omega(s)=\sqrt{(\beta/a)|\alpha_{x}v_{x}^{2}(s)+\alpha_{y}v_{y}^{2}(s)+\frac{3}{2}(\alpha_{x}+\alpha_{y}-4\alpha_{z})v_{z}^{2}(s)|}$
is non-zero for all $s$. The anisotropic case $\alpha_{x}=\alpha_{y}\gg\alpha_{z}$
is shown in Figure 3c. The phase space of parameters with $\alpha_{x}=\alpha_{y}$
that leads to a zero-field QHE is shown in the light red areas of
Fig. 3d.

The deformation pattern $\mathbf{u}=(2xz,2yz,0)$ can be created with
the strain forces indicated by the arrows in Fig. 4a. Interestingly,
the physical implementation of the family of deformations (\ref{eq:u-1})
can be achieved with the application of a uniform temperature gradient
along the $z$ axis of the crystal (see Fig. 4b). Since $\mathbf{u}$
describes the displacement of the lattice sites from their equilibrium
position, the thermal expansion is represented as $u_{i}=\Delta x_{i}=x_{i}\gamma_{i}\Delta T\propto\gamma_{i}x_{i}z$,
where $\gamma_{i}=\text{d}x_{i}/\text{d}T$ is the linear thermal
expansion coefficient in the $i=x,y,z$ direction and $\Delta T(z)=T-T_{0}\propto z$
is the temperature variation from equilibrium. This tunable pattern
of deformations could be created with temperature gradients in crystals
and acoustic metamaterials \cite{zhu-1}.

\begin{figure}[t]
\begin{centering}
\includegraphics[scale=0.35]{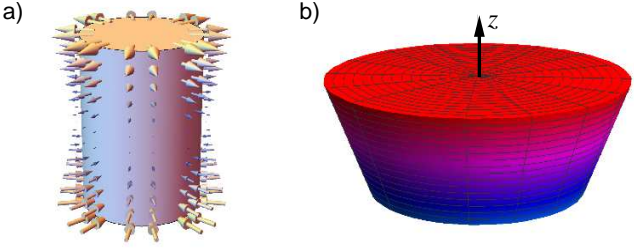}\vspace{-0.2cm}
\par\end{centering}
\caption{{\small{}(a) Elastic deformation of a cylinder under the strain configuration
$\mathbf{u}=(2xz,2yz,0)$. The arrows indicate the strain forces that
create uniform nearly flat LLs in a 3D material (see fig. 3c). (b)
Temperature gradient along the $z$ axis that implements the strain
field $\mathbf{u}=(\alpha_{x}xz,\alpha_{y}yz,\alpha_{z}z^{2})$, with
$\alpha_{i}$ ($i=x,y,z$) proportional to the thermal expansion coefficients.
Red: hot region. Blue: cold.}}
\end{figure}

\emph{Elastic Hall viscosity.} In quantum Hall systems, the Hall viscosity
follows from the linear response of the system to gravitational fluctuations,
which manifest through local changes in the metric of space $\xi_{ij}=\frac{1}{2}(\partial_{i}\xi_{j}+\partial_{j}\xi_{i})$,
where $\xi_{i}$ has the physical meaning of a strain field. The so
called gravitational Hall viscosity is defined as the variation of
the stress tensor $T_{\mu\nu}=\partial\mathcal{H}/\partial\xi_{\mu\nu}$
to time variations of the strain tensor $\dot{\xi}_{ij}$. By analogy,
the elastic (phonon) Hall viscosity can be derived using linear response
as \cite{Barkeshli,Shapourian,Avron1,Avron2}
\begin{equation}
\left\langle \frac{\partial\mathcal{H}_{p}}{\partial u_{\mu\nu}}\right\rangle =\lambda_{\mu\nu\rho\gamma}u_{\rho\gamma}+\eta_{\mu\nu\rho\gamma}\dot{u}_{\rho\gamma}
\end{equation}
where $\langle\ldots\rangle$ integrates over the fermions, $\lambda_{\mu\nu\rho\gamma}$
is the elastic moduli, $\dot{u}_{\rho\gamma}$ the strain-rate tensor
and $\eta_{\mu\nu\rho\gamma}$ the elastic Hall viscosity tensor.
The first term is the elastic response of a charge neutral fluid and
the second one the viscous response \cite{Avron1,Avron2}. As the
stress tensor, the tensors $u$, $\dot{u}$ are symmetric, while the
viscosity tensor is symmetric under $\mu\leftrightarrow\nu$ or $\rho\leftrightarrow\gamma$.
However, with respect to the exchange $\mu\nu\leftrightarrow\rho\gamma$,
the viscosity tensor has a symmetric part $\eta_{\mu\nu\rho\gamma}^{S}=\eta_{\rho\gamma\mu\nu}^{S}$
and an antisymmetric one $\eta_{\mu\nu\rho\gamma}^{A}=-\eta_{\rho\gamma\mu\nu}^{A}$.
The symmetric part is associated with dissipation and vanishes at
zero temperature. The antisymmetric one describes a non-dissipative
response with topological nature and is non-zero only when TRS is
broken. In general, one can calculate the antisymmetric viscosity
tensor from the effective action 
\begin{equation}
\delta S_{H}=\frac{1}{2}\int d^{3}x\,dt\,\eta_{\mu\nu\rho\gamma}u_{\mu\nu}\dot{u}_{\rho\gamma},\label{eq:SH}
\end{equation}
which resembles a Chern-Simons action for the usual QHE \cite{Hughes,Cortijo}. 

We will consider the elastic Hall viscosity for the 3D QAH state,
which is an extension of the Haldane model for the hyperhoneycomb
lattice, described in detail in ref. \cite{Kim}. For nodal line semimetals,
loop currents on the lattice can create a mass term around the nodal
line with the general form 
\begin{equation}
\mathcal{H}_{m}(\mathbf{q})=\left[m(s)+\!\!\sum_{i=x,y,z}v_{i}^{\prime}(s)q_{i}\right]\sigma_{3},\label{Hm}
\end{equation}
where $v_{i}^{\prime}(s)$ is gives the mass dispersion in the $i=x,y,z$
direction. The Haldane mass $m(s)$ changes sign at $2(2n+1)$ points
along the nodal line, with $n\in\mathbb{N}$, breaking inversion and
TRS symmetry \cite{Kim,Okugawa}. The nodes of the mass, where $m(s)=0$,
are Weyl points with a well defined helicity \cite{Kim}. Weyl points
with opposite helicities are connected by surface states in the form
of topological Fermi arcs \cite{Armitage}. 

\emph{Effective action. }In the QAH state, the Hamiltonian away from
the Weyl points of the nodal line has the form 
\begin{equation}
\mathcal{H}_{\textrm{QAH}}(\mathbf{q})=\mathcal{H}_{p}(\mathbf{q})+m(s)\sigma_{3}.\label{eq:HQAH}
\end{equation}
The effective action in terms of the strain tensor $u_{ij}$ can be
derived by integrating out the fermions. That results in the effective
action $S_{\textrm{ef}}(u)=\textrm{Tr}\left[\ln\left(G^{-1}\right)\right]$,
where $G^{-1}(q)=iq_{0}-\mathcal{H}_{QAH}(\mathbf{q})\equiv G_{0}^{-1}(q)-\Sigma_{el}$
is the Green's function and 
\begin{equation}
\Sigma_{el}(u)=v_{z}A_{1}\sigma_{1}+(v_{x}A_{2}+v_{y}A_{2}^{\prime})\sigma_{2}\label{SE}
\end{equation}
is the self-energy due to elastic terms. For convenience, we defined
the elastic gauge fields in (\ref{eq:el}) as $A_{1}=-\frac{\beta}{a}\frac{3}{4}(u_{xx}+u_{yy}-2u_{zz})$,
$A_{2}=-\frac{\beta}{a}u_{xz}$ and $A_{2}^{\prime}=-\frac{\beta}{a}u_{yz}$. 

Expanding the action in powers of the elastic gauge fields, namely
$S_{\textrm{ef}}=\textrm{tr}\ln G_{0}^{-1}-\text{tr}\sum_{n=0}^{\infty}\frac{1}{n}\left(G_{0}\Sigma\right)^{n}$,
the lowest order contribution to the Hall viscosity comes from two
loop, $S_{\textrm{ef}}^{(2)}=-\frac{1}{2}\textrm{tr}\left[G_{0}\Sigma G_{0}\Sigma\right]$.
More explicitly, 
\begin{align}
\delta S_{\textrm{ef}} & =-\frac{1}{2}\int\!\frac{d^{4}k}{\left(2\pi\right)^{4}}\left[v_{x}v_{z}A_{1}(-k)\Pi^{12}(k)A_{2}(k)\right.\nonumber \\
 & \qquad\left.+v_{y}v_{z}A_{1}(-k)\Pi^{12}(k)A_{2}^{\prime}(k)+(1\leftrightarrow2)\right],\label{eq:deltaS2}
\end{align}
where $\Pi^{\mu\nu}(k)=\int\frac{d^{4}q}{\left(2\pi\right)^{4}}\textrm{tr}\left[G_{0}\left(q+k\right)\sigma_{\mu}G_{0}\left(q\right)\sigma_{\nu}\right]$
is the standard polarization tensor, with antisymmetric off-diagonal
terms, $\Pi^{12}(k)=-\Pi^{21}(k)$. Integration can be done by slicing
the BZ into planes intersecting the nodal line at two points. Integrating
over a slice in the $xz$ plane for the first term,
\begin{equation}
v_{x}v_{z}\Pi^{12}(k)=\frac{k_{0}}{2\pi}\int_{C}\frac{dq_{y}}{\left(2\pi\right)}\nu_{(y)}(\mathbf{q}_{0})=-\frac{k_{0}}{2\pi}\lambda_{y},\label{Pi12}
\end{equation}
where $\nu_{(y)}(\mathbf{k}_{0})=\frac{1}{2}\text{sign}[v_{x}(s)v_{z}(s)m(s)]=\pm\frac{1}{2}$
is the topological charge of 2D massive Dirac fermions confined to
an $xz$ plane crossing the nodal line at $\mathbf{k}_{0}$. Integration
along the nodal loop $C$ gives the $y$ component of the Chern vector
$\boldsymbol{\lambda}=(\lambda_{x},\lambda_{y},\lambda_{z})$, which
is belongs to the reciprocal lattice $\mathbf{G}$ and sets the 3D
quantum Hall conductivity of the system, $\sigma_{ij}=e^{2}/(2\pi h)\epsilon_{ijk}\lambda_{k}$
. From a similar argument, $v_{y}v_{z}\Pi^{12}(k)=-v_{y}v_{z}\Pi^{21}(k)=k_{0}\lambda_{x}/2\pi$.
Hence, 
\begin{align}
\delta S_{\textrm{ef}} & =\frac{1}{16\pi^{2}}\int\!\frac{d^{4}k}{\left(2\pi\right)^{4}}\left[-\lambda_{x}A_{1}(-k)k_{0}A_{2}^{\prime}(k)\right.\nonumber \\
 & \qquad\qquad\left.+\lambda_{y}A_{1}(-k)k_{0}A_{2}(k)-(1\leftrightarrow2)\right].\label{eq:deltaS1}
\end{align}

Performing the substitution $A_{1}=d_{z}$, $A_{2}=d_{x}$ and $A_{2}^{\prime}=d_{y}$,
the effective action can be written in a more compact form,
\begin{equation}
\delta S_{\textrm{ef}}=\frac{1}{16\pi^{2}}\int d^{4}x\,\epsilon^{\mu\nu\rho}\lambda_{\mu}d_{\nu}\dot{d_{\rho}},\label{deltaS3}
\end{equation}
where
\begin{align}
d_{x} & =-\frac{\beta}{a}u_{xz},\nonumber \\
d_{y} & =-\frac{\beta}{a}u_{yz}\label{dy}\\
d_{z} & =-\frac{\beta}{a}\frac{3}{4}\left(u_{xx}+u_{yy}-2u_{zz}\right).\nonumber 
\end{align}
 For the hyperhoneycomb lattice, the Chern vector is $\boldsymbol{\lambda}=\mathbf{b}_{1}+\mathbf{b}_{2}=(2\pi/\sqrt{3},-2\pi/\sqrt{3},0)a^{-1}$
\cite{Kim}. Writing the action in a more explicit form,
\begin{align}
\delta S_{\textrm{ef}} & =\frac{1}{2}\int d^{4}x\,\eta_{H}\left[(u_{xx}+u_{yy}-2u_{zz})(\dot{u}_{yz}+\dot{u}_{xz})\right.\nonumber \\
 & \qquad\qquad\left.-(u_{yz}+u_{xz})(\dot{u}_{xx}+\dot{u}_{yy}-2\dot{u}_{zz})\right],\label{eq:DeltaSu}
\end{align}
with $\eta_{H}=\beta^{2}\sqrt{3}/8\pi a^{3}$. The action can be cast
in the form of (\ref{eq:SH}), where the elastic Hall viscosity tensor
is $\eta_{xxxz}=\eta_{xxyz}=\eta_{yyxz}=\eta_{yyyz}=\eta_{H},$ and
$\eta_{zzxz}=\eta_{zzyz}=-2\eta_{H}$. The elastic Hall viscosity
tensor is anisotropic, as expected in 3D \cite{Avron2}, and reflects
the topological nature of the QAH state \cite{note}. In nodal-line
semimetals, the Chern vector is related to the arclength separating
two Weyls points \emph{along} the nodal line. Hence, the shape of
the nodal line contains information about the lattice and can be used
even in effective low energy models to determine the \emph{exact}
elastic Hall viscosity in terms of the elastic parameter and lattice
constant $a$.

\emph{Experimental observation.} Although there are no known examples
of semimetallic hyperhoneycomb crystals \cite{Modic}, this lattice
may be artificially created in optical lattices \cite{Jotzu}, and
also in photonic \cite{Rechtsman,Lu} and acoustic metamaterials \cite{zhu-1}.
In twisted graphene bilayers, elastic gauge fields can be created
with electric field effects \cite{Ramires}. In synthetic lattices,
strain deformations can be readily implemented with local displacements
of the lattice sites, without the need to apply pressure. While local
probes such as scanning tunneling spectroscopy can fully characterize
the LLs in 2D \cite{Levy,Gomes}, this method can be used to characterize
the surface states of the LLs in the 3D case.

In quantum Hall systems, the measurement of the Hall viscosity is
typically challenging \cite{Berdyugin}, as it involves probing the
response of the stress tensor under changes of the space metric \cite{Avron2}.
In Galilean invariant systems in the hydrodynamic regime, the Hall
viscosity can be determined solely in terms of the electromagnetic
response due to a non-homogeneous electric field \cite{Hoyos,Haldane2}.
The elastic Hall viscosity nevertheless can be measured in terms of
the dispersion of sound waves. When $\eta_{H}=\eta_{xxxz}$ is zero,
the longitudinal and transverse modes are decoupled at long wavelengths.
In the topological phase, where $\eta_{H}$ is finite, the transverse
and longitudinal modes are expected to mix, allowing one to measure
the elastic Hall viscosity through the corrections to the dispersion
of the phonons \cite{Barkeshli}. The quantum simulation of Chern
insulating phases has been done in honeycomb lattices of cold atoms
\cite{Jotzu}, in quantum circuits \cite{Roushan} and acoustic metamaterials
\cite{zhu-1}. We conjecture that the QAH state in 3D may be experimentally
realized in synthetic lattices as well.

\emph{Conclusions. }We have derived the elastic gauge fields that
are created due to lattice deformations in the hyperhoneycomb lattice.
We proposed a family of strain configurations that lead to uniform
nearly flat LLs in 3D. The strain fields can be created with the application
of uniform temperature gradient, driving a controllable reconstruction
of the bulk states into nearly flat LLs. That raises the prospect
of engineering tunable zero-field 3D QHE in metamaterials. In the
topological phase, we have also shown that the components of the elastic
Hall viscosity tensor in the 3D QAH state for this lattice are $\pm\eta_{H}$
or $\pm2\eta_{H}$ (or zero), with $\eta_{H}=\beta^{2}\sqrt{3}/\left(8\pi a^{3}\right)$. 

\emph{Acknowledgements. }SWK thanks X. Dou for helpful discussions.
BU and SWK acknowledge NSF CAREER grant No DMR-1352604 for support.

\end{document}